# A MULTI-RESOLUTION FRONT-END FOR END-TO-END SPEECH ANTI-SPOOFING

*Wei Liu[1], Meng Sun[1*], Xiongwei Zhang[1], Hugo Van hamme[2], Thomas Fang Zheng[3]*

[1] Lab of Intelligent Information Processing, PLAAEU, Nanjing, China
[2] KU Leuven, PSI, Dept. of Electrical engineering (ESAT), Leuven, Belgium
[3] Center for Speech and Language Technologies, BNRist, Tsinghua University, Beijing, China

**ABSTRACT**

The choice of an optimal time-frequency resolution is usually a difficult but important step in tasks involving speech signal classification, e.g., speech anti-spoofing. The variations of the performance with different choices of time-frequency resolutions can be as large as those with different model architectures, which makes it difficult to judge what the improvement actually comes from when a new network architecture is invented and introduced as the classifier. In this paper, we propose a multi-resolution front-end for feature extraction in an end-to-end classification framework. Optimal weighted combinations of multiple time-frequency resolutions will be learned automatically given the objective of a classification task. Features extracted with different time-frequency resolutions are weighted and concatenated as inputs to the successive networks, where the weights are predicted by a learnable neural network inspired by the weighting block in squeeze-and-excitation networks (SENet). Furthermore, the refinement of the chosen time-frequency resolutions is investigated by pruning the ones with relatively low importance, which reduces the complexity and size of the model. The proposed method is evaluated on the tasks of speech anti-spoofing in ASVSpoof 2019 and its superiority has been justified by comparing with similar baselines.

## 1. INTRODUCTION

The choice of an optimal time-frequency resolution is usually a difficult job in the tasks involving speech signal classification and recognition. Ad hoc window length and frame shift are usually taken in automatic speech recognition (ASR) or automatic speaker verification (ASV), which is partially based on the prior knowledge of phonetics and also considers the stationarity required by Fourier transforms. However, in some classification tasks where it is not clear what cues in which scales contribute much to the final performance of the classification, we are not always lucky to have chosen a relatively good time-frequency resolution in the experiments. Speech anti-spoofing is a good example of this kind of tasks, which is an important step to secure an ASV system but still lacks interpretability on how time-frequency features work to classify genuine utterances from spoofing ones.

Existing techniques for speech spoofing include, but are not limited to, record-and-replay, text-to-speech (TTS) and voice conversion (VC). Those techniques generate voices to pass the verification stage of an ASV system by claiming that the voices come from a target present speaker. Anti-spoofing is therefore to prevent these techniques from spoofing an ASV system [1]. The ways how spoofing utterances are created (i.e., recorded or synthesized) will definitely have a great impact on how anti-spoofing works, which include the various silence-voice intervals in record-and-replay and the miscellaneous time-frequency scales utilized in waveform generation in TTS and VC. In ASVspoof2019, both TTS and VC are used to create voices for spoofing purpose in the logical access (LA) track [2]. For example, a 25.5ms window length and a 5ms frame shift are taken by the WaveNet vocoder to generate speech samples in [3], a combination of 50ms window length and 12.5ms frame shift is used in the end-to-end TTS in [4], the voice conversion in [5] adopts a 25ms window length and a 5ms frame shift, while a 20ms window length and a 10ms frame shift is taken in [6]. This implies a single time-frequency resolution in an anti-spoofing model may not be sufficient to cope with so many choices of window lengths and frame shifts when synthesizing voices.

Researchers have tried to introduce structures with multiple resolutions or scales to improve the neural network classifier itself. For example, convolutional blocks with different scales are used to process the same input time-frequency features for end-to-end automatic speech recognition in [7-10]. A channel wise gating mechanism is introduced to enforce the connections between feature groups in neural networks in [11]. A channel attention module is introduced to convolutional neural networks (CNN) where the key features of anti-spoofing are explored by assigning weights to different positions and channels in a feature map [12]. However, those works are based on time-frequency features with a fixed resolution. Though multiple scales in the configuration of network structures can capture information across different scales, some detailed information may have been lost in the early stage of feature extraction.

Actually, a predetermined time-frequency resolution may not necessarily be the most suitable choice for the backend neural network classifier. Furthermore, the variations of the performance with different time-frequency resolutions can be as large as those with different neural network architectures, which makes it difficult to judge what the performance gain actually comes from when a new network architecture is invented and evaluated as the backend classifier.

Towards extracting features with multiple time-frequency resolutions, multi-resolution feature maps are investigated in CNN for speech anti-spoofing in a recent work [13]. Time-frequency features with multiple window lengths are first extracted offline and then stacked as inputs with different channels to CNN. In this paper, we improve the work in [13] by 1) imposing learnable weights to the multi-resolution features, 2) optimizing them in an end-to-end framework, and 3) also considering the variations of frame shifts (i.e., besides the variations of window lengths).

In summary, this paper proposes a multi-resolution plug-in learnable front-end for speech anti-spoofing. In Section 2, we will present how to optimize the weighted combinations of features with different time-frequency resolutions. The refinement of the chosen time-frequency resolutions is also investigated by pruning the ones with relatively low importance. Experimental settings, results and analysis are given in Section 3 and 4. The conclusion is in Section 5.

## 2. THE PROPOSED METHOD

The proposed multi-resolution front-end is illustrated in Figure 1 where its role in the classification task using SENet is also depicted. The proposed method consists of four parts: differentiable feature extraction using a Short Time Fourier Transformation (STFT) layer, an adaptive pooling layer to align time-frequency features with different sizes, a learnable block to predict weights, and SENet as the backend classifier. An additional branch, *refinement by pruning resolutions with low weights*, is also proposed and investigated to reduce the complexity and size of the model.

### 2.1. A STFT layer to promote end-to-end learning

A STFT layer realized by *torch.stft* computes the Fourier transform of short overlapping windows of the input waveforms for each configuration of the time-frequency resolution. The log energy spectrum of the features with each resolution is subsequently computed by applying $log(|\cdot|)$ on the complex spectrum from *torch.stft*, and denoted as a feature map $W_m \times H_m$ in Figure 1. Compared to the feature extraction procedures in a preprocessing step (e.g., using Kaldi or librosa to extract and save STFT features), there operations make the overall learning in an end-to-end fashion, which will ease the subsequent weights prediction in Section 2.3 and the refinement of the combinations of resolutions in Section 2.5.

### 2.2. Adaptive pooling to algin and stack feature maps

Feature maps extracted with different resolutions have different sizes, so alignment is required before stacking them. An adaptive pooling layer is adopted to perform *upsampling* to automatically align the feature maps, where each feature map is reshaped to have the maximal scales of all the feature map, denoted by $W_{max} \times H_{max}$ in Figure 1. The comparison of upsampling with other alignment methods will be discussed later in Section 4.3.

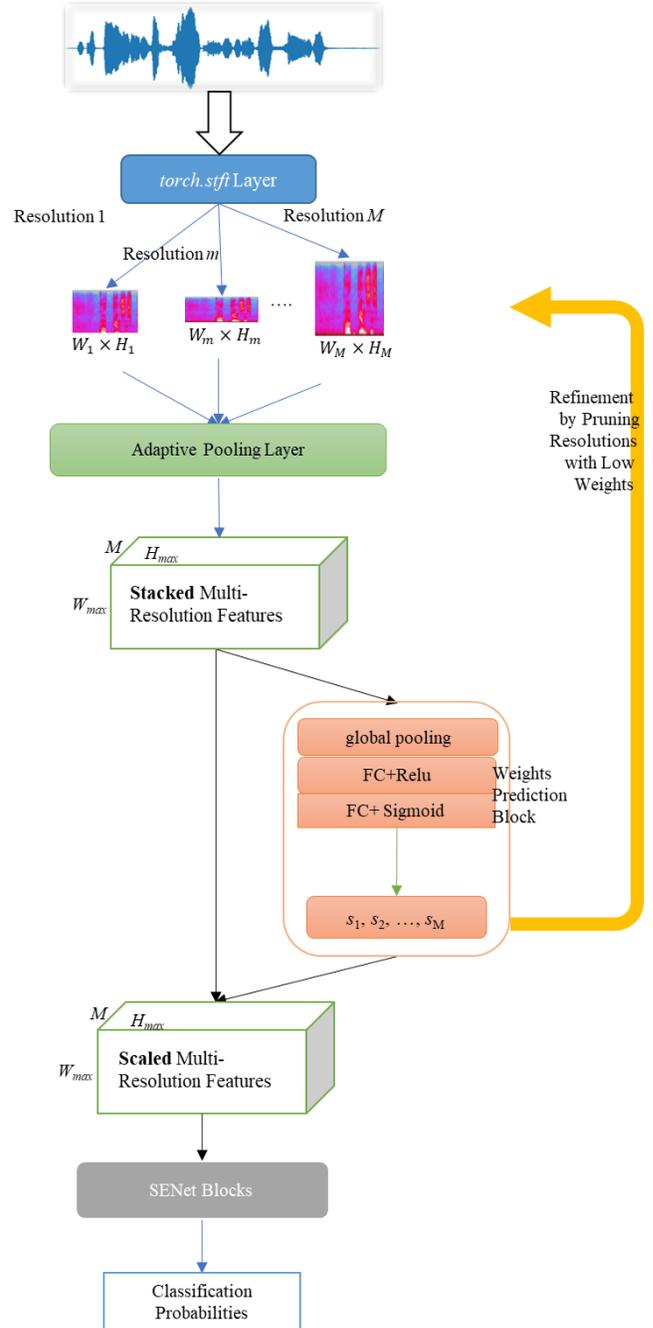

Figure 1. The proposed multi-resolution front-end and its role in a speech classification task using SENet.

## 2.3. Weight prediction to scale features

The weight prediction is modeled by a learnable neural network inspired by the Squeeze-and-Excitation (SE) block from [14]. As denoted in the middle part of Figure 1, the stacked multi-resolution feature maps with size $M \times W_{max} \times H_{max}$ first go through *global pooling* to form a $M \times 1 \times 1$ vector by averaging the entries of the feature map with each resolution.

Therefore, each two-dimensional feature map is condensed into a real number representing the full receptive field. Two Fully Connected (FC) layers with ReLU and Sigmoid activations are subsequently used as the regressor to predict the weights of the feature maps, see $s_1, s_2, \ldots, s_M$ in Figure 1. The stacked feature maps are then scaled by the weights as inputs to the following SENet classifier.

## 2.4. SENet for speech anti-spoofing

SENet is known for its focus on the channel relationship and has demonstrated great superiority on tasks of both image classification [14] and speech anti-spoofing [15]. It holds the ability to adaptively recalibrate channel-wise feature maps by explicitly modelling interdependencies between channels. In this paper, we adopt SENet34 as the backend classifier, please refer to [15] for more details.

## 2.5. Pruning feature maps with low weights

When implementing the multi-resolution recipe presented above, redundant features which contribute little to the final classification task could be introduced. After finishing the training with $M$ resolutions, we can use the learned weights to perform pruning. We sort the learned weights in an ascending order, compute the differences of any two adjacent values, and take the index of the maximal one. That is,

$$m^* = \arg\max_{m} \{s_{m+1} - s_m, m = 1, \cdots, M\} \quad (1)$$

where $s_m$'s represent the sorted weights. The last $m^*$ resolutions will be retained as important ones. As will be observed in Section 4.1, the refinement will not only reduce the model's complexity but could also improve the model's performance.

## 3. EXPERIMENTAL SETTINGS

### 3.1. Datasets and Metrics

The dataset used to evaluate the proposed method and the relevant baselines comes from the challenge of ASVspoof 2019 [2] by following the evaluation plan strictly [16]. Both equal error rates (EER) and the minimum normalized tandem detection cost function (t-DCF )[17] are reported.

### 3.2. Network Configuration

The network structure of SENet34 is taken the same as that in [15] which consists of 16 residual blocks from [18] and their corresponding Squeeze-and-Excitation blocks to assign weights to each channel.

### 3.3. Training Details

All the utterances were unified to have the same length of 4.5 seconds. For the ones shorter than 4.5s, repetition of the utterance was conducted to make extension; while for the ones longer than 4.5s, random segments of 4.5s were chosen. Actually, most of the utterances in the dataset are shorter than 4.5s.

We followed the optimization recipe presented in [15], where *adam* was used to optimize the model with $\beta_1 = 0.9$, $\beta_2 = 0.98$, and the weight attenuation $10^{-9}$. In the first 1000 warm-up steps, a learning rate planner was used to increase the learning rate linearly, and then the learning rate was reduced gradually according to the reciprocal of the square root of the number of steps [19]. At each training epoch, the best model was retained based on its performance on the *dev* part of the data.

## 4. RESULTS AND ANALYSIS

### 4.1. Performance of SENet with Multiple Choices of Resolutions

In one's naïve intuition, the candidate resolutions should include as many choices as possible, in case of missing important ones beneficial to the back-end task. Therefore, we consider 3 window lengths (512, 1024, and 2048) and 4 frame shifts (64, 128, 256, and 512) and their combinations listed in Table 1. Four additional choices are also included by considering their good performance reported in literatures, e.g., the window/frame combinations 400/160 from [15] and 1724/130 from [20] with outstanding results on ASVSpoof 2019-LA, and the combinations 288/96 and 480/120 both from [13] with outstanding results on ASVSpoof 2019-PA.

Table 1. The results of single resolution and multi-resolution models.

| | Resolution (window/shift) | LA | | PA | |
|---|---|---|---|---|---|
| | | EER | t-DCF | EER | t-DCF |
| Hand Selected Resolution | 512 / 64 | 10.60 | 0.270 | 4.06 | 0.095 |
| | 512 /128 | 11.56 | 0.240 | 3.40 | 0.075 |
| | 1024 / 64 | 16.72 | 0.409 | 2.89 | 0.069 |
| | 1024 /128 | 8.15 | 0.210 | 2.33 | 0.055 |
| | 1024 /256 | 11.72 | 0.192 | 2.76 | 0.062 |
| | 2048 / 64 | 5.72 | 0.165 | 1.79 | 0.042 |
| | 2048 /128 | 9.88 | 0.261 | 2.20 | 0.050 |
| | 2048 /256 | 4.67 | 0.147 | 2.53 | 0.056 |
| | 2048 /512 | 5.54 | 0.107 | 2.69 | 0.057 |
| | 400 /160 [15] | 9.24 | 0.200 | 3.75 | 0.082 |
| | 1724 /130 [22] | 5.38 | **0.098** | 2.10 | 0.050 |
| | 288 / 96 [13] | - | - | 3.91 | 0.091 |
| | 480 /120 [13] | - | - | 3.09 | 0.073 |
| | Top-3 | 10.21 | 0.242 | 2.66 | 0.060 |
| Proposed | Full | 5.43 | 0.135 | **1.07** | **0.024** |
| | Refined | **3.67** | 0.110 | 1.24 | 0.025 |

All the choices above are denoted by "Hand Selected Resolution" in Table 1. From the table, we found that window length plays a more important role than frame shift where longer windows tend to perform better than shorter ones, on both LA and PA. 1724/130 seems a lucky oracle choice holding good performance but with a relatively short window. However, we may not be always so lucky for new tasks.

The "Full" row in Table 1 represents the results of combining all the resolutions above where the combination is conducted by adaptive pooling. Good results have been observed by outperforming all the single resolutions in PA and approximating the performance of the single-best resolution in LA.

### 4.2. Refinement of Multi-Resolution Combinations

Multi-resolution combinations can be further refined by removing the ones with low weights, according to the proposed approach in Section 2.5. The weights are depicted in Figure 2, where the refined resolutions are 1024/256, 1724/130, and 512/128 for LA and 2048/64, 480/120, and 1024/64 for PA, as denoted in the "Refined" row of Table 1. In order to evaluate the effectiveness of the proposed refinement method, combinations with three resolutions holding the top performance were selected for comparison purpose. For example, according to Table 1, the top-3 best resolutions for LA are 2048/256, 1724/130, and 2048/512, while the ones for PA are 2048/64, 1724/130, and 2048/128. The results are reported in the "Top-3" row of Table 2, which are inferior to our proposed method "Refined".

From Table 1, we observe that the refinement cannot only reduce the model's complexity by avoiding extracting features of resolutions with low weights, but could also improve the performance of the "Full" combination, e.g., the reduce of EER from 5.43 to 3.67 for LA.

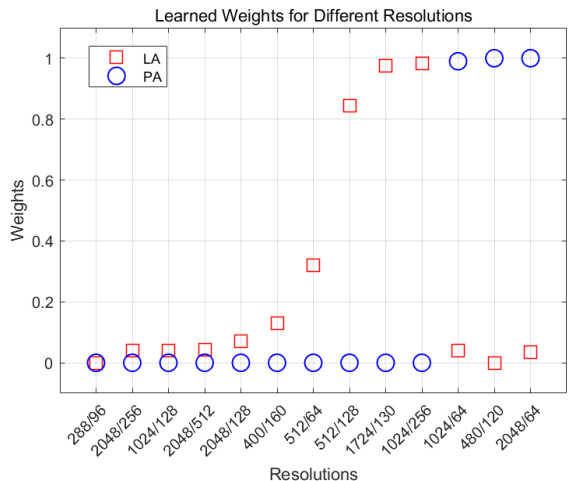

Figure 2. The learned weights for different time-frequency resolutions.

### 4.3. Comparison on the Alignments of Feature Maps

As presented in Section 2.2, we have chosen the adaptive pooling to algin feature maps with different sizes. However, different ways of alignment have different impacts on the model's performance. In this section, besides adaptive pooling, we evaluate two alternative choices of the alignment of feature maps: upsampling from [21] and deconvolution from [22]. Three resolutions, 512/128, 1024/256, and 2048/256, are chosen in the experiments. The results in Table 2 demonstrate the superiority of adaptive pooling.

Table 2. Performance of the proposed methods with three different alignment methods on LA.

| Alignment | EER | t-DCF |
|---|---|---|
| **Adaptive Pooling** | **4.47** | **0.121** |
| Upsample | 5.11 | 0.124 |
| Deconvolution | 7.12 | 0.198 |

### 5. RELATION TO PRIOR WORK

The most similar work to our paper is [13], where an offline multi-resolution feature fusion method to perform speech anti-spoofing was studied in the PA dataset of ASVspoof 2019. Three different window lengths of 18ms, 25ms and 30ms were chosen for evaluation. However, they did not consider the variations of the importance of different resolutions. We improve [13] in three aspects: 1) imposing learnable weights on the multi-resolution features, 2) optimizing them in an end-to-end framework, and 3) also considering the variations of frame shifts (i.e., besides the variations of window lengths). Since SeNet50 was used in [13], we conducted an additional group of experiments on PA by using SeNet50 to make intuitive comparisons. The results are shown in Table 3 where superiority of our method was observed.

Table 3. Comparison of our multi-resolution front-end and the one from [13] on PA

| Model | SENet50 | | SENet34 | |
|---|---|---|---|---|
|  | EER | t-DCF | EER | t-DCF |
| Multi-Res. In [13] | 1.76 | 0.040 | - | - |
| Ours | **1.53** | **0.038** | **1.24** | **0.025** |

### 6. CONCLUSION

In speech signal classification tasks, the choice of time-frequency resolutions has a great impact on the model's performance even using the same backend model structure. In this paper, we have proposed a multi-resolution front end for feature extraction in an end-to-end classification framework.

The front-end is able to automatically exploit the best weighted combinations of resolutions to fit well to the classification task. By pruning the resolutions with low weights, the model's complexity is reduced and its performance can be improved. The proposed method has demonstrated its superiority on both the PA and LA tracks of the ASVspoof 2019 dataset. It is worth noting that our method can be integrated into any kind of neural networks as a front-end and can be applied to any classification task of speech signals.